\documentclass[sigconf]{acmart}
\usepackage{subcaption}
\usepackage{graphicx}
\usepackage{transparent}
\usepackage{xcolor}

\copyrightyear{2025}
\acmYear{2025}
\setcopyright{acmlicensed}
\acmConference[ICMR '25] {Proceedings of the 48th International ACM ICMR Conference on Research and Development in Information Retrieval}{ June 30--July 3, 2025}{Chicago, IL, USA.}
\acmBooktitle{Proceedings of the 48th International ACM ICMR Conference on Research and Development in Information Retrieval (ICMR '25), June 30--July 3, 2025, Chicago, IL, USA}
\acmISBN{979-8-4007-1592-1/25/07}
\acmDOI{10.1145/3731715.3733461}

\begin{document}

\title{ViFusion: In-Network Tensor Fusion for Scalable Video Feature Indexing}

\begin{abstract}

Large-scale video feature indexing in datacenters is critically dependent on efficient data transfer. Although in-network computation has emerged as a compelling strategy for accelerating feature extraction and reducing overhead in distributed multimedia systems, harnessing advanced networking resources at both the switch and host levels remains a formidable challenge. These difficulties are compounded by heterogeneous hardware, diverse application requirements, and complex multipath topologies. Existing methods focus primarily on optimizing inference for large neural network models using specialized collective communication libraries, which often face performance degradation in network congestion scenarios.

To overcome these limitations, we present ViFusion, a communication aware tensor fusion framework that streamlines distributed video indexing by merging numerous small feature tensors into consolidated and more manageable units. By integrating an in-network computation module and a dedicated tensor fusion mechanism within datacenter environments, ViFusion substantially improves the efficiency of video feature indexing workflows. The deployment results show that ViFusion improves the throughput of the video retrieval system by 8--22× with the same level of latency as state-of-the-art systems.

\end{abstract}



%   \label{fig:arch}



\author{Yisu Wang}
\affiliation{%
  \institution{The Hong Kong University of Science and Technology (Guangzhou)}
   \department{Information Hub}
  \city{Guangzhou}
  \country{China}
}
\email{ywang418@connect.hkust-gz.edu.cn}

\author{Yixiang Zhu}
\affiliation{%
  \institution{The Hong Kong University of Science and Technology (Guangzhou)}
  \department{Information Hub}
  \city{Guangzhou}
  \country{China}
}
\email{yixiangzhu@hkust-gz.edu.cn}

\author{Xinjiao Li}
\affiliation{%
  \institution{The Hong Kong University of Science and Technology (Guangzhou)}
  \department{Information Hub}
  \city{Guangzhou}
  \country{China}
}
\email{xinjiaoli@hkust-gz.edu.cn}

\author{Yulong Zhang}
\affiliation{%
  \institution{The Hong Kong University of Science and Technology (Guangzhou)}
  \department{Information Hub}
  \city{Guangzhou}
  \country{China}
}
\email{yzhang893@connect.hkust-gz.edu.cn}

\author{Ruilong Wu}
\affiliation{%
  \institution{The Hong Kong University of Science and Technology (Guangzhou)}
  \department{Information Hub}
  \city{Guangzhou}
  \country{China}
}
\email{rwu408@connect.hkust-gz.edu.cn}

\author{Dirk Kutscher}
\authornote{Corresponding author.}
\affiliation{%
  \institution{The Hong Kong University of Science and Technology (Guangzhou)}
  \department{Information Hub}
  \city{Guangzhou}
  \country{China}
}
\email{dku@hkust-gz.edu.cn}

% \author{Anonymous Authors}
% \affiliation{%
%   % \institution{Paper under Double-Blind Review}
%   \country{}
% }

\maketitle

\begin{figure*}[!t]
  % \hspace{-25px}  % 调整这个值来控制偏左的程度
  \includegraphics[scale=2]{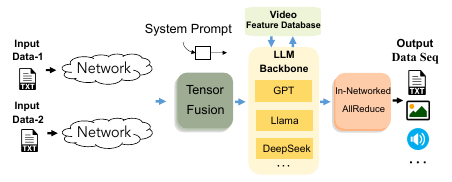}
  \caption{ViFusion Processing Pipeline. Multimodal video features are processed through a combination of tensor fusion, LLM inference, and in-network optimization. }
  \label{fig:fig1}
\end{figure*}

\section{Introduction}

% \dirk{suggestion for the intro: keep this rather short and move some of the motivation text and explanation to section 2}
The rapid expansion of video-centric digital services has placed tremendous pressure on large-scale feature indexing systems to efficiently process increasingly complex queries. Modern video understanding models, from temporal action detection frameworks \cite{he2022asmlocactionawaresegmentmodeling} to multimodal architectures ~\cite{wadekar2024evolutionmultimodalmodelarchitectures}, require high computational throughput and extensive memory bandwidth\cite{vivit, Diffusion}. Despite significant advances in GPU acceleration, single-node deployment remains impractical for models with billions of parameters\cite{openai_sora_2024, song2020generativemodelingestimatinggradients}, which necessitates distributed execution across clusters.

Compared with traditional document or image retrieval, video retrieval presents unique and significant challenges. First, there is an inherent semantic gap between video and text. Text is highly abstract symbolic information, whereas video comprises sequences of visual frames rich in spatio-temporal signals. These differences in representation and semantic distribution make cross-modal alignment particularly difficult. Second, the massive amount of video data introduces substantial computational overhead and communication costs. Retrieving a specific segment from a long video may involve decoding and analyzing tens of thousands of frames. When processing such data in a distributed setting, additional overhead arises from data transmission and synchronization across nodes, which can significantly impact system throughput and response latency. Thus, efficient handling of long videos while achieving precise cross-modal alignment and temporal grounding is a core challenge in video retrieval.

% Second, videos possess a strong temporal dimension: a text query typically corresponds to an event within a specific time segment. This requires systems to have robust temporal grounding capabilities to locate relevant segments accurately within lengthy videos. In contrast to static image-text matching, video retrieval must capture and align dynamic, multiframe visual semantics. For instance, a query may involve a continuous sequence of human actions or causal event relationships.

To address these issues, researchers are exploring large-scale pre-trained models to improve video retrieval \cite{Diffusion, vivit}. Traditional methods relying on manual tags or rules cannot fully capture video semantics or user intent. In contrast, multimodal pre-trained models and LLMs can bridge the video-text gap by leveraging representations learned from massive data. For example, CLIP \cite{clip} aligns images and text via contrastive learning, while GPT-4 \cite{gpt4} and DeepSeek-R1 \cite{deepseek} demonstrate strong language understanding and reasoning. Integrating these models into video indexing goes beyond simple keyword matching and more accurately captures the deeper semantics of user queries.

However, deploying large models directly in video-retrieval services is challenging. The state-of-the-art LLMs are extremely large: DeepSeek-R1 \cite{deepseek}, for example, contains 671 billion parameters, requiring the model to be partitioned between multiple GPUs during both training and inference. Similarly, large multimodal models used for video understanding are often too large to fit or be inferred efficiently on a single GPU \cite{guo2024cephaloharnessingheterogeneousgpu, he2024malmmmemoryaugmentedlargemultimodal, deepseed-moe}. Moreover, multi-machine cooperative inference introduces imbalances in computing and communication resources. Differences in computational speed and network bandwidth among nodes can lead to additional synchronization overhead, and the cost of communication may offset the benefits of parallelism, resulting in limited throughput gains or even degraded performance with increased system latency \cite{cachegen}. 

% \dirk{do we have a reference for these statements?}

In this paper, we propose ViFusion, a system that addresses these challenges by combining batch-based inference with in-network computation. We fuse video feature queries into a single large tensor and offload partial computations to in-network computation nodes.

Figure ~\ref{fig:fig1} illustrates the architecture of our proposed system, which integrates tensor fusion and in-network optimization to enhance video retrieval performance. Multiple text-based inputs are received via the network and aggregated through the tensor fusion module. The fused representation is processed by the LLM backbone and enriched with relevant video features from the video feature database. To enhance efficiency, the system employs in-network AllReduce for optimized communication. Finally, the processed data is output as videos, providing multimodal retrieval results.

By dynamically batching incoming requests and fusing them into unified tensors, ViFusion maximizes the throughput of deep learning accelerators, reduces per-query memory overhead, and amortizes inference costs across multiple queries. Even small requests are processed as part of a larger fused batch, significantly improving GPU utilization and overall system throughput.

Beyond batching optimization, in-network aggregation can accelerate distributed feature indexing in ViFusion. Instead of performing all tensor aggregation on end-host GPUs, part of the reduction process is offloaded to middle host computation node. This in-transit aggregation minimizes communication overhead, alleviates network congestion, and improves scalability, while maintaining correct results within network hardware constraints. In addition, ViFusion seamlessly integrates with popular deep learning frameworks, such as PyTorch\cite{pytorch}.

% Our ViFusion implementation aims to support popular DNN training frameworks, including PyTorch \cite{pytorch}, with APIs and native interfaces similar to those of BytePS \cite{byteps} and Horovod \cite{Horovod}.

The contributions of this study are as follows.

\begin{itemize}
\item A communication-aware tensor fusion framework that jointly optimizes batch sizes, network utilization, and GPU occupancy for video indexing workloads.

\item An in-network computation strategy that offloads and optimizes AllReduce operations to boost overall performance.

\item Comprehensive experimental validation demonstrating that, compared to existing methods, our approach achieves significant improvements in both retrieval accuracy and system efficiency, validating the effectiveness and practical value of the proposed system.

\end{itemize}

% \dirk{the intro is quite long and not easy to follow. Maybe re-factor this and mention only selected things in the intro?}

\section{Related Work and Motivation}

\subsection{LLM for Video Search Integration}

Researchers are currently exploring the application of LLMs in video retrieval to enhance the semantic capabilities of video indexing and understanding. One approach is to leverage multimodal models to connect textual queries with video content. For example, OpenAI’s CLIP model \cite{clip}, which was originally designed for image-text tasks, has been extended to video retrieval. A representative method, CLIP4Clip \cite{clip4clip}, utilizes CLIP’s visual-semantic representations to map video frame features into a shared embedding space with text, enabling text-to-video retrieval. By transferring CLIP’s knowledge to video-language tasks, CLIP4Clip achieved state-of-the-art performance on multiple video retrieval benchmarks.

Beyond embedding-based matching, another approach is to use LLMs as video-understanding engines. Multimodal LLMs such as GPT-4V \cite{gpt-4v} and LLaVA \cite{liu2023visualinstructiontuning} can analyze images, making them suitable for generating semantic descriptions of video content. 

GPTSee \cite{gptsee} is an example of such a two-stage framework. In the first stage, it applies MiniGPT-4 \cite{minigpt-4} to generate descriptions for the video frames and rewrites the query. In the second stage, it computes the semantic similarity between the descriptions and the query to locate relevant video segments. Similarly, Goldfish \cite{Goldfish} introduced a retrieval-augmented generation framework for long-video question-answering tasks. It first uses MiniGPT4-Video \cite{MiniGPT4-Video} to generate rich descriptions for video segments, and then retrieves relevant segments based on the query for LLMs to process, enabling effective comprehension of videos of arbitrary length.

Despite these advances, existing LLMs still face notable challenges in video search. Chief among these is limited temporal modeling and constrained context windows. Most multimodal LLMs, including GPT-4V \cite{gpt-4v}, are unable to directly ingest lengthy videos because of their finite context size. LLaVA, for example, can only handle around two to three frames, and while GPT-4V is more powerful, it remains bounded by context-length constraints when attempting to capture a video’s global context. This shortcoming substantially hinders performance in fine-grained retrieval tasks like pinpointing specific temporal segments.

Overcoming these bottlenecks often demands scaling video analytics across distributed clusters to handle massive or continuous video streams. However, distributed systems bring their own complexities, particularly regarding internode communication \cite{cachegen}. Frequent exchanges of intermediate results and parameters are typically required, relying on collective communication (CC) \cite{nccl} operations such as AllReduce for synchronizing updates and aggregating features \cite{pytorch}.

Another key challenge is the inefficient handling of query batching. Traditional systems process each video query independently, triggering redundant inference passes and data transfers \cite{pytorch, tensorflow}. Although batched inference can significantly improve GPU throughput by amortizing model loading and maximizing parallelism\cite{vllm, clipper}, asynchronous query arrivals and heterogeneous workloads often force production systems to use suboptimal batch sizes, exacerbating communication overhead and underutilizing expensive hardware.

Previous video indexing systems have not fully resolved these challenges. Specialized models \cite{avs} may gain speed but lose generalizability, while offline batch processing \cite{batch-query} can introduce unacceptable delays for time-sensitive applications. We contend that a unified approach is necessary, which is one that co-optimizes tensor fusion, network utilization, and batching strategies to meet real-time requirements.

ViFusion, which focuses on integrating LLMs with video processing, can leverage CLIP4Clip’s cross-modal embedding approach for text-video alignment, where LLMs generate video descriptions for semantic matching. By combining LLMs' powerful language understanding with dedicated video feature extraction and retrieval modules, ViFusion has the potential to achieve breakthroughs in complex video semantic queries.

\subsection{In-Network Computation in Data Centers}

In-Network Computation (INC) \cite{switchml} has recently gained attention for data center acceleration by leveraging programmable switches, SmartNICs, and other network devices such as FPGA to perform partial computations during data transmission \cite{inc-survey}. This approach shifts certain application-layer tasks to the network layer, thereby reducing communication overhead and boosting system throughput for distributed tasks, such as deep learning and large-scale video processing.

Traditionally, network protocols have enforced a strict separation between host-based computation and stateless forwarding in the network layer, ensuring line-rate performance. However, the surge in the demand for distributed deep learning has turned communication into a critical bottleneck. Systems such as SwitchML \cite{switchml}, ATP \cite{atp}, netRPC \cite{NetRPC}, and ClickINC \cite{clickinc} demonstrate how programmable switches such as Intel’s P4 Tofino \cite{p4} and Trio \cite{trio} can serve as in-network aggregators to accelerate collective operations (e.g., AllReduce), thereby effectively reducing the overhead associated with multinode synchronization.

Many of these solutions assume a simple single-switch topology and have not been extended to multi-rack architectures common in large-scale deployments of frameworks, such as BytePS \cite{byteps} and Horovod \cite{Horovod}, where hierarchical, multi-tier datacenter networks cause substantial inter-rack or inter-datacenter traffic. As a result, existing INC methods struggle to handle complex topologies in which the randomized allocation of GPU/CPU resources leads to increased physical hop counts, higher congestion, and longer communication times. 

However, by offloading computations closer to the data, INC offers a new dimension of performance optimization for systems. For instance, Planter \cite{planter} provides a modular and efficient open-source framework for rapid prototyping of in-network machine learning models across a range of platforms and pipeline architectures. 

\begin{figure*}[!t]
  \centering
  \includegraphics[scale=0.3]{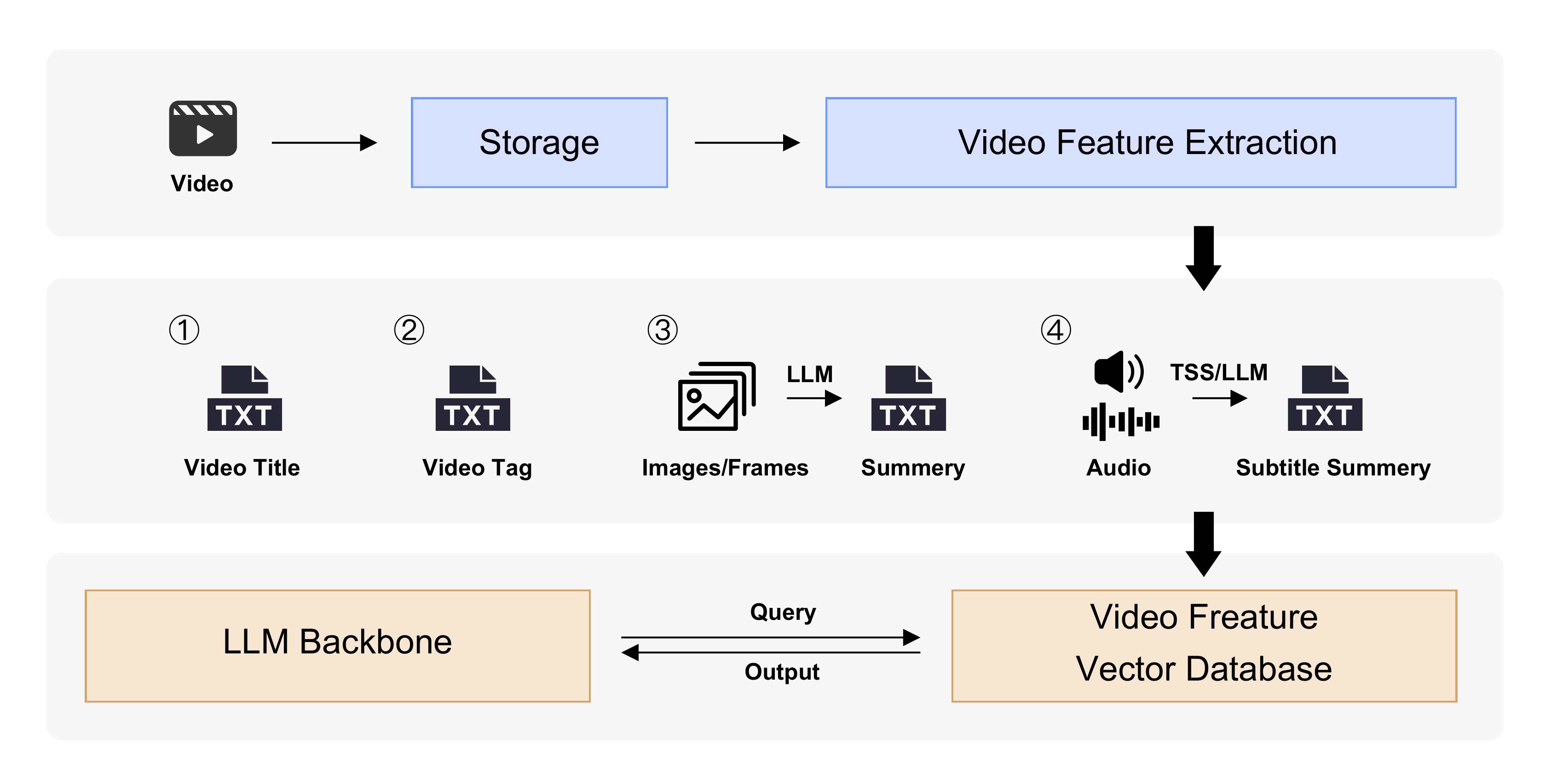}
  \caption{ViFusion Multi-Modal Video Feature Extraction and Retrieval. The process begins with the ingestion of raw video into a Storage module, followed by Video Feature Extraction, where multimodal representations are generated. A LLM Backbone interacts with the database to process user queries, retrieving relevant video content based on semantic understanding of the indexed video features.}
  \label{fig:overview}
\end{figure*}

% 

% \subsection{Large Language Model Inference Optimization}

% \begin{figure*}[htbp]
%     \centering
%     \begin{minipage}[b]{0.32\textwidth}
%         \centering
%         \includegraphics[width=\textwidth, scale=0.25]{asset/group INA.pdf}
%         \caption{Example of ViFusion In-Network AllReduce Module Architecture}
%         \label{fig:arch}
%     \end{minipage}
%     \hfill
%     % \begin{minipage}[b]{0.32\textwidth}
%     %     \centering
%     %     \includegraphics[width=\textwidth, scale=1]{asset/INC arch.pdf}
%     %     \caption{In-network aggregation example}
%     %     \label{fig:inc topo}
%     % \end{minipage}
%     % \hfill
%     % \begin{minipage}[b]{0.32\textwidth}
%     %     \centering
%     %     \includegraphics[width=\textwidth, scale=1]{asset/datacenter.pdf}
%     %     \caption{A typical inter-datacenter network}
%     %     \label{fig:datacenter}
%     % \end{minipage}
% \end{figure*}

\section{SYSTEM DESIGN}

ViFusion’s architecture is further detailed in Figure~\ref{fig:overview}, and is designed to efficiently handle multimodal video content by unifying storage, feature extraction, and indexing within a single framework. The pipeline begins with the ingestion of raw video into a storage module, followed by the generation of rich multimodal representations that include textual metadata, visual frames, and audio features. Video titles and tags provide structured descriptors, key frames are extracted and summarized via a LLM, and audio is transcribed before being similarly distilled. These outputs are consolidated and stored in the Video Feature Vector Database such as Milvus \cite{Milvus}, forming the basis for subsequent retrieval. When a query arrives, the LLM Backbone taps into this database to retrieve semantically relevant content, ensuring accurate and scalable search results. To further enhance performance, Tensor Fusion unifies heterogeneous input streams into a single representation with system prompts for contextual guidance, whereas in-network AllReduce efficiently coordinates distributed computations. Together, these components enable ViFusion to produce text, images, and audio outputs at scale, significantly improving indexing efficiency and retrieval speed. 

\subsection{Tensor Fusion Module}

\begin{figure}[!t]
  \centering
  \includegraphics[scale=0.3]{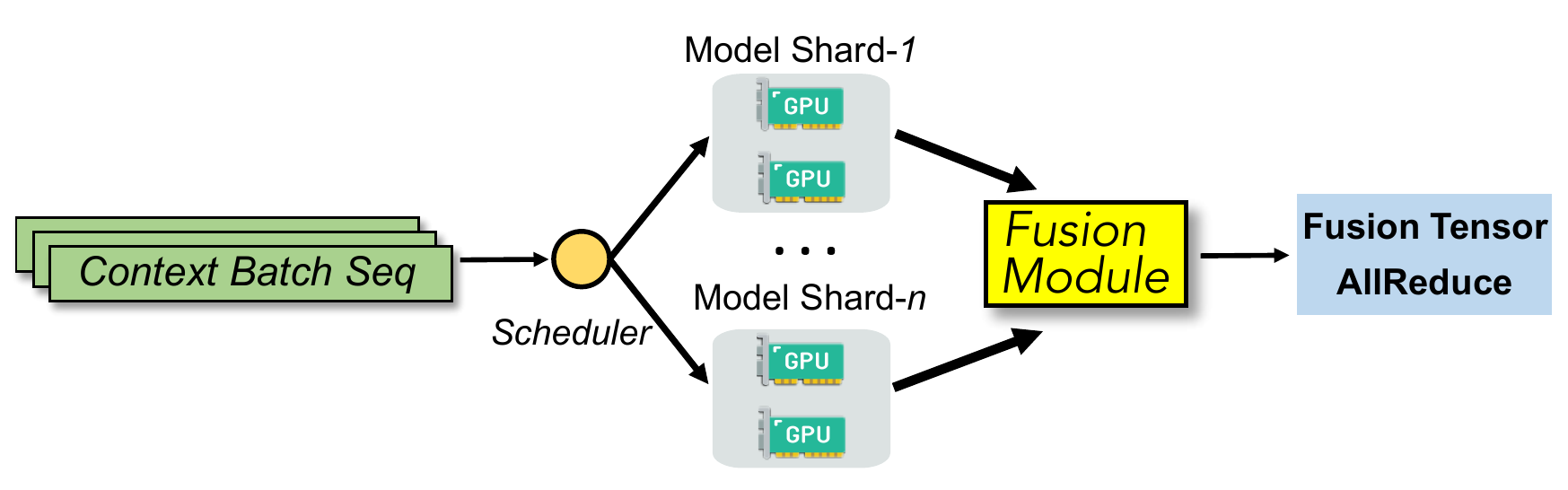}
  \caption{Example of Tensor Fusion Module}
  \label{fig:tensor fusion}
\end{figure}

Efficient processing of video queries in large-scale clusters requires addressing two key bottlenecks: redundant parameter transfers from per-request execution and poor hardware utilization when small clips fail to saturate modern accelerators. As illustrated in Fig.~\ref{fig:tensor fusion}, ViFusion’s \textit{Tensor Fusion Module} (TFM) addresses these issues by dynamically grouping incoming video segments into larger fused tensors, distributing their computation across multiple GPUs, and carefully orchestrating data transfers to minimize the overhead. This section describes the core TFM design principles, including multilevel model sharding, zero-copy tensor concatenation, and a scheduler-driven fusion mechanism tailored to real-time video workloads.

TFM begins by treating incoming queries as temporally coherent units, rather than as independent tasks. A central buffer manager places arriving frames into pinned GPU memory regions, grouped by modalities such as RGB or optical flow. To accommodate real-time constraints while preserving high throughput, the system employs a \emph{dual-threshold triggering} policy. Specifically, it initiates fusion operations when (1) the combined data in the GPU buffer reaches approximately 75\% of the capacity or (2) a 150 ms deadline elapses. Under low arrival rates, batches are formed promptly to maintain responsiveness; under bursts of traffic, TFM fuses small segments together to better amortize data transfers and kernel launches.

Another key architectural feature of TFM is \textit{ multilevel model sharding}. Rather than replicating the entire model on each device or limiting sharding to two large partitions, TFM supports partitioning the network into multiple shards (e.g., ModelShard-1 to ModelShard-n), distributing consecutive layers among different GPUs. Thus, each GPU holds only the parameters for its assigned shard and exchanges only intermediate activation tensors. Early partition boundaries (shallow splits) balance compute load but can generate larger activation transfers if the earlier layers produce high-dimensional features. Conversely, deeper splits yield smaller activation tensors but potentially skew the workload if one shard dominates the model depth. The TFM policy engine can profile layer-level compute and activation sizes to select split points that minimize communication overhead and maintain similar processing times across GPUs. During pipeline execution, the output of each shard is immediately transferred to the GPU, which begins processing as soon as it is free, thereby overlapping communication and computing.

In addition to careful partitioning, TFM reduces memory-copy overhead with \emph{zero-copy concatenation}. Traditional approaches often copy partial outputs to CPU memory for merging and then transfer the fused data back to the GPU, thereby incurring redundant data movement and high latency. Instead, TFM relies on direct GPU-to-GPU communication (e.g., GPUDirect RDMA), enabling one GPU to write its intermediate results into a designated buffer on another GPU without traversing the host memory. In the context of multimodal or multi-segment inference, this means that the optical flow and RGB features can be concatenated in a unified address space via peer-to-peer DMA, eliminating CPU involvement. This zero-copy mechanism is crucial for sustaining a high throughput when repeatedly merging small tensors in streaming workloads.

TFM also incorporates a strategy called \textit{partial temporal alignment} to optimize memory layouts. Notably, most state-of-the-art video models require inputs padded to fixed lengths to accommodate tensor-friendly operations within deep learning frameworks. Instead of padding all sequences to match the length of the longest clip in a batch, TFM takes a more nuanced approach by padding each sequence to a power-of-two size. This decision capitalizes on low-level hardware optimizations: memory alignment is more efficient at powers-of-two, reducing computational overhead when reading and writing data in GPU memory. Additionally, such alignment often enables better utilization of kernel operations in libraries like cuDNN or cuBLAS, where thread block configurations and boundary checks are simplified.

% Beyond merging, TFM also integrates \textit{partial temporal alignment} to optimize memory layouts. Many state-of-the-art video models expect inputs padded to fixed lengths for tensor-friendly operation. Rather than padding all segments to the longest clip in a batch, TFM pads sequences to a power of $2$ sizes. The system then tracks the actual spatiotemporal shape of each query such that only valid portions are processed at subsequent stages, preserving correctness while avoiding wasted GPU cycles.

These fused and aligned tensors flow through a \textit{scheduler-based iteration loop}, designed to avoid head-of-line blocking. Once a fused batch is dispatched to a pipeline shard, the TFM immediately accepts new arrivals in the pinned buffer and attempts to schedule them if resources are available. When a shard finishes processing, the scheduler checks for any backlog of waiting segments and assigns them to the now-free GPU shard. This continuous loop overlaps data transfers with kernel execution; for example, the final DMA writes from one shard occurring concurrently with kernel launches on other shards. Moreover, TFM’s memory manager employs fixed-size pool allocations on the GPU to avoid fragmentation, coupled with a spillover mechanism that can temporarily stage data in host memory for sudden bursts.

Figure~\ref{fig:tensor fusion} shows how each query flows from right to left through the pipeline. Newly arrived queries join pinned buffers on the right side, are partially aligned, and are concatenated across modalities. Once a batch crosses the threshold, it enters the pipeline. Topology-aware scheduling can then prioritize intra-rack bandwidth and forward data along high-speed links to minimize inter-rack congestion. After processing, the completed queries exit on the left side, freeing resources for subsequent batches. This design extends fine-grained batching concepts from large-scale inference systems \cite{vllm, distserve} to accommodate the peculiarities of video analytics such as asynchronous arrival patterns, multimodal inputs, and high-dimensional spatiotemporal data.

\subsection{in-network AllReduce}

\begin{figure}[!t]
  \centering
  \includegraphics[scale=0.2]{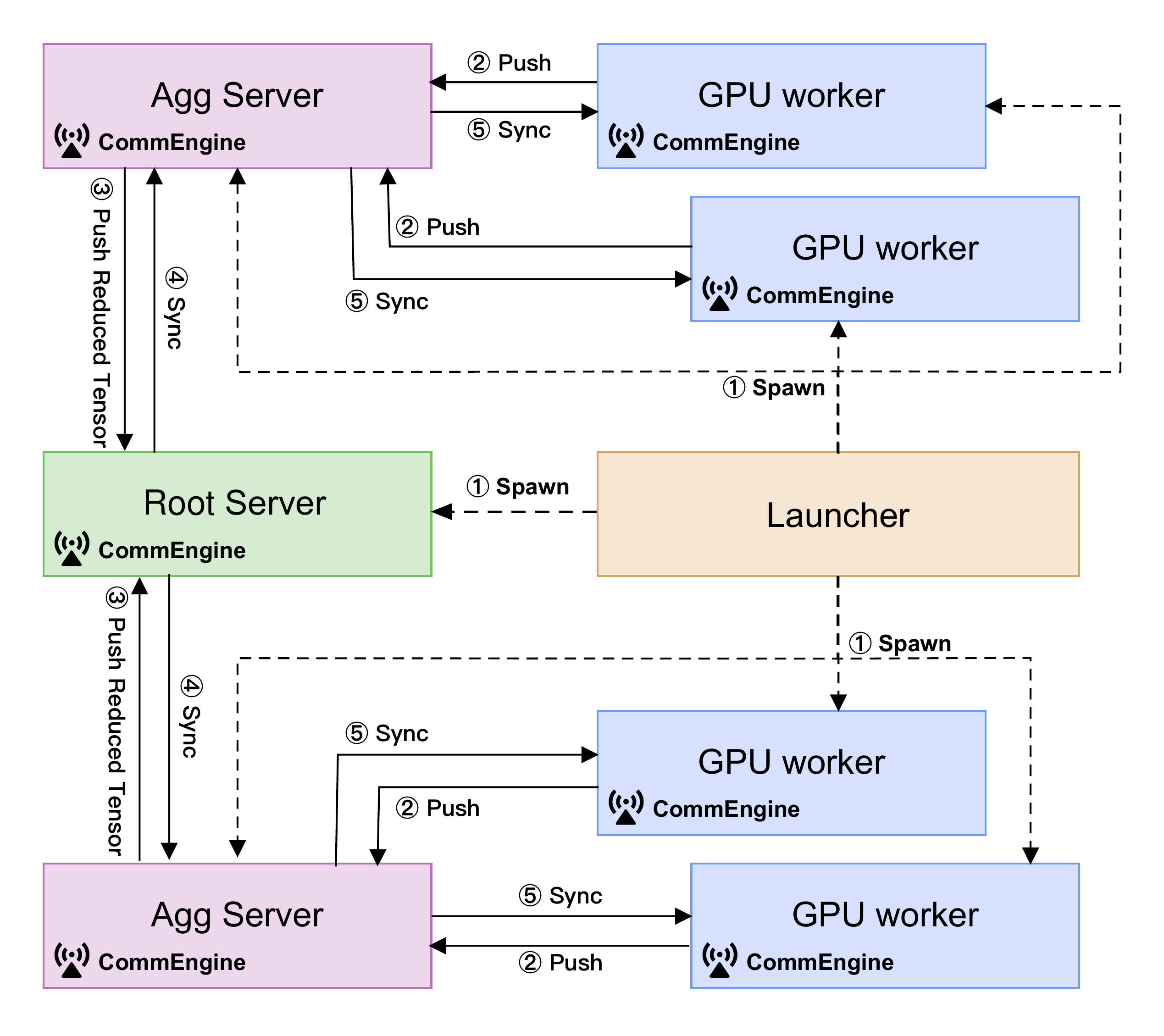}
  \caption{In-Network AllReduce Execution Flow}
  \label{fig:ina exe}
\end{figure}

In ViFusion’s LLM backbone, tensor-parallel inference is adopted, in which large model layers are distributed across multiple devices. Each device computes a partial result that must be aggregated to form the final output. For instance, in a row-partitioned linear layer, each GPU holds a partial product, requiring an AllReduce operation to sum across the devices. Unlike training, inference lacks a backward pass to hide communication overhead, making any AllReduce in the forward path a direct contributor to end-to-end latency. While distributed training can mask much of the network communication through overlapping computation, inference places these communication interactions on the critical path, making their optimization essential for low-latency performance \cite{gherghescu2024ivegot99problems}.

\begin{figure}[!t]
  \centering
  \includegraphics[scale=0.35]{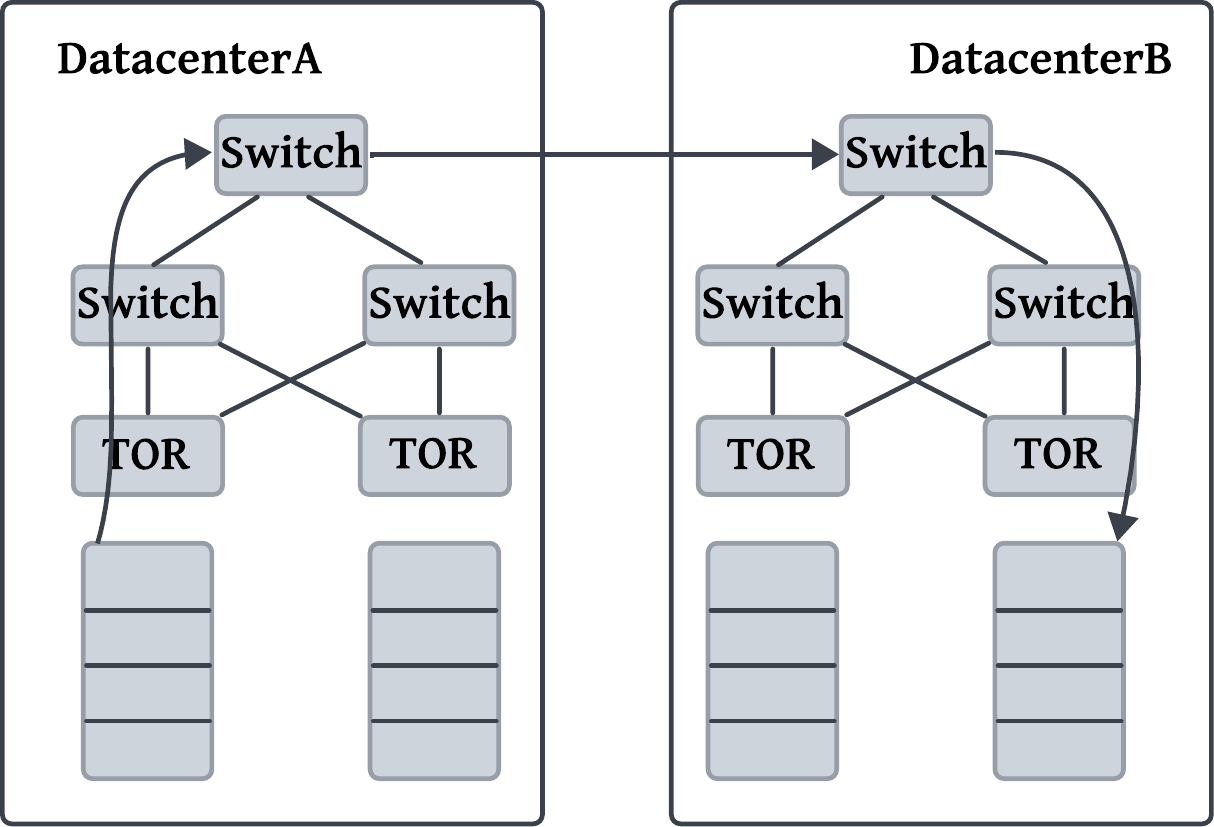}
  \caption{A typical inter-datacenter network}
  \label{fig:datacenter}
\end{figure}

During inference, the primary bottleneck in AllReduce arises from the communication imbalance caused by network topology. As shown in Figure \ref{fig:datacenter}, GPUs in different clusters are distributed across multiple racks, interconnected via a multi-tier network. Intra-rack bandwidth is typically high, whereas inter-rack links often suffer from limited bandwidth or oversubscription. When AllReduce spans multiple racks, large data transfers across congested inter-rack links can introduce significant delays.

Traditional AllReduce methods such as single-ring AllReduce across all nodes treat all connections equally, leading to potential bottlenecks at the slowest link. For example, an overloaded uplink on a core switch can be a performance constraint. Excessive inter-rack traffic can severely degrade AllReduce throughput, turning the network uplink into a bottleneck \cite{rat}. This imbalance results in stragglers, where certain nodes are forced to wait for slower communication, ultimately degrading the overall inference time.

\begin{figure}[!t]
  \centering
  \includegraphics[scale=0.2]{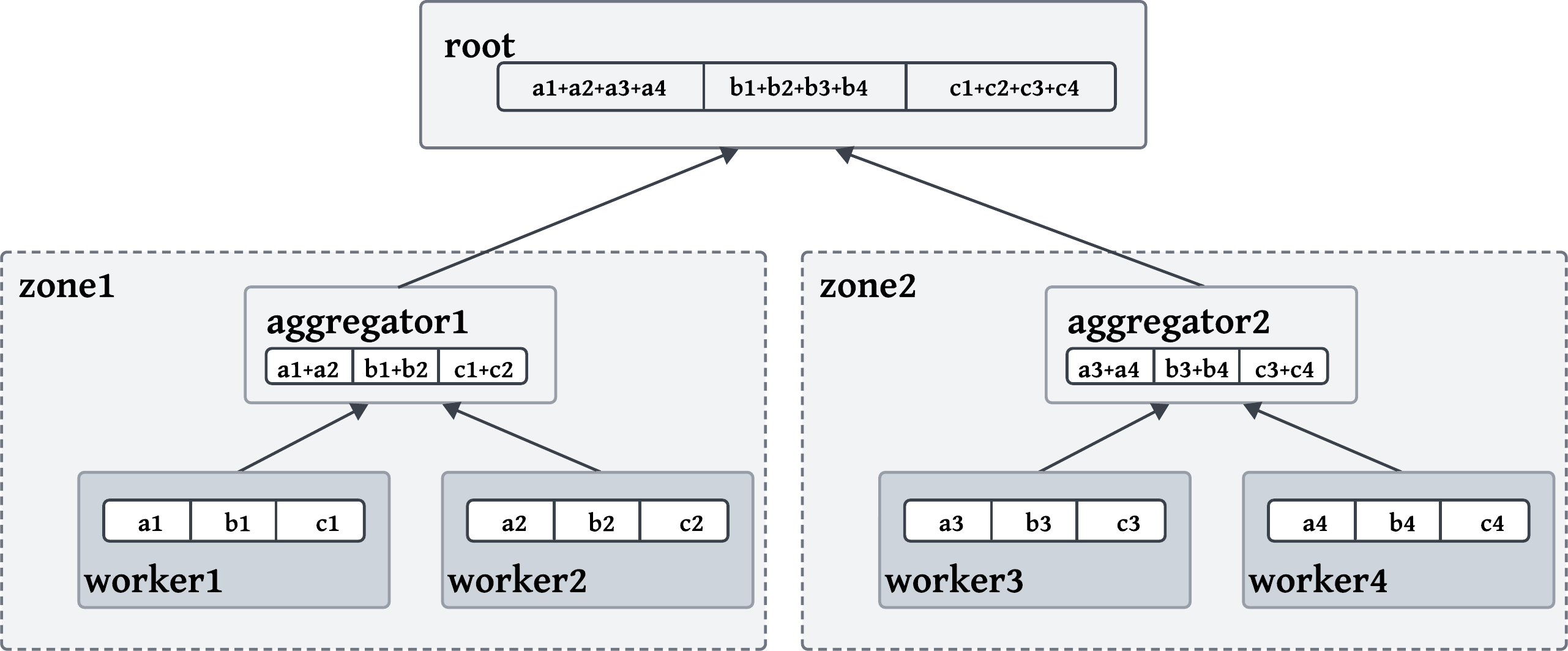}
  \caption{Example of ViFusion In-Network AllReduce Module Architecture}
  \label{fig:ina_arch}
\end{figure}

To address these bottlenecks, ViFusion employs an In-Network AllReduce mechanism optimized for inference. Figure \ref{fig:ina_arch} illustrates the architecture of the ViFusion In-Network AllReduce module, highlighting its key features of group-wise aggregation and server-based execution. 

We introduce a hierarchical AllReduce scheme aligned with the physical network topology to reduce costly cross-rack communication. The key idea is to perform reductions in a two-tier manner: first within local groups (e.g., within each rack or server pod) and then across groups. All GPUs are partitioned into clusters based on proximity. For instance, all GPUs under the same top-of-rack switch form a single group. An AllReduce operation is first executed inside each group, quickly combining data over the high-bandwidth, low-latency local links. Subsequently, a second-stage AllReduce is performed among the group leaders (or a designated aggregator per group) to combine the partial aggregates into a global result. This design dramatically reduces the volume of data that must traverse inter-rack links; instead of every node communicating across the rack boundary, only one message per group needs to go through. Hierarchical schemes are known to alleviate network imbalance by confining most of the traffic to local domains.

In ViFusion’s case, after the intra-rack AllReduce, the cross-rack AllReduce involves far fewer participants and bytes, which avoids saturation of the uplinks. This approach is analogous to the multilevel reductions used in large-scale data-parallel training (e.g., first within a rack, then across racks), but here it is applied to inference tensor-parallel communication. By matching the reduction hierarchy to the network hierarchy, we ensure that slower links are used sparingly.

As shown in Figure \ref{fig:ina exe}, the execution diagram illustrates how ViFusion leverages In-Network AllReduce within inference systems. The process begins with a launcher that spawns GPU workers across different nodes to effectively manage job distribution. Once operational, each GPU worker computes the intermediate tensors, pushing them to aggregation servers for hierarchical reductions. These servers use INC (within dedicated servers) to minimize data transfer. After intermediate aggregation, synchronization steps ensure that the aggregated results are consistently distributed back to the GPU workers, thus preserving model coherence.

Although originally designed for distributed training, this In-Network AllReduce model is readily adaptable for inference acceleration. In inference-heavy scenarios, particularly for distributed video feature-indexing tasks, the system aggregates feature tensors rather than gradients. By integrating ViFusion’s tensor fusion and AllReduce operations, small tensors are batched into larger tensors prior to aggregation, reducing both communication overhead and inter-node latency.

% {\bfseries Your document will be returned to you for revision if
%   modifications are discovered.}

\section{Evaluation}

\begin{figure*}[!t]
  \centering
  % 子图 (a)
  \begin{subfigure}[b]{0.45\textwidth}
    \centering
    \includegraphics[scale=0.45]{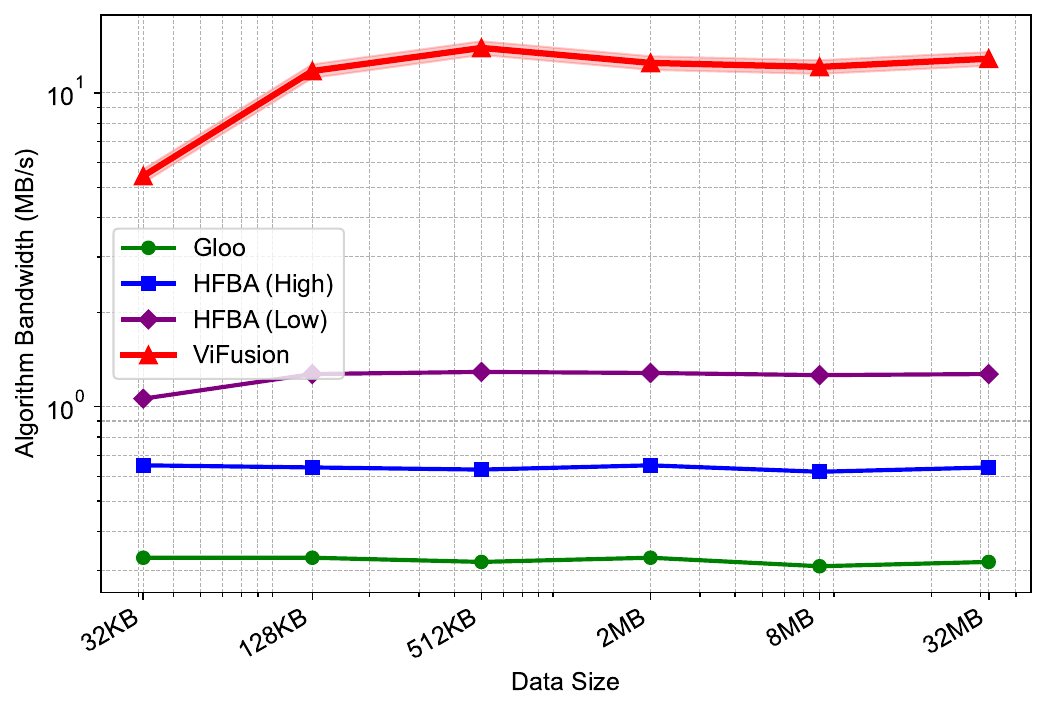}
    \caption{Comparison of the bandwidth of AllReduce in 4 GPU workers}
    \label{fig:4gpu_allreduce_ina}
  \end{subfigure}
  \hfill
  % 子图 (b)
  \begin{subfigure}[b]{0.45\textwidth}
    \centering
    % 如果有8 GPU的图片，则请替换文件名，例如：asset/8GPU_allreduce.pdf
    \includegraphics[scale=0.45]{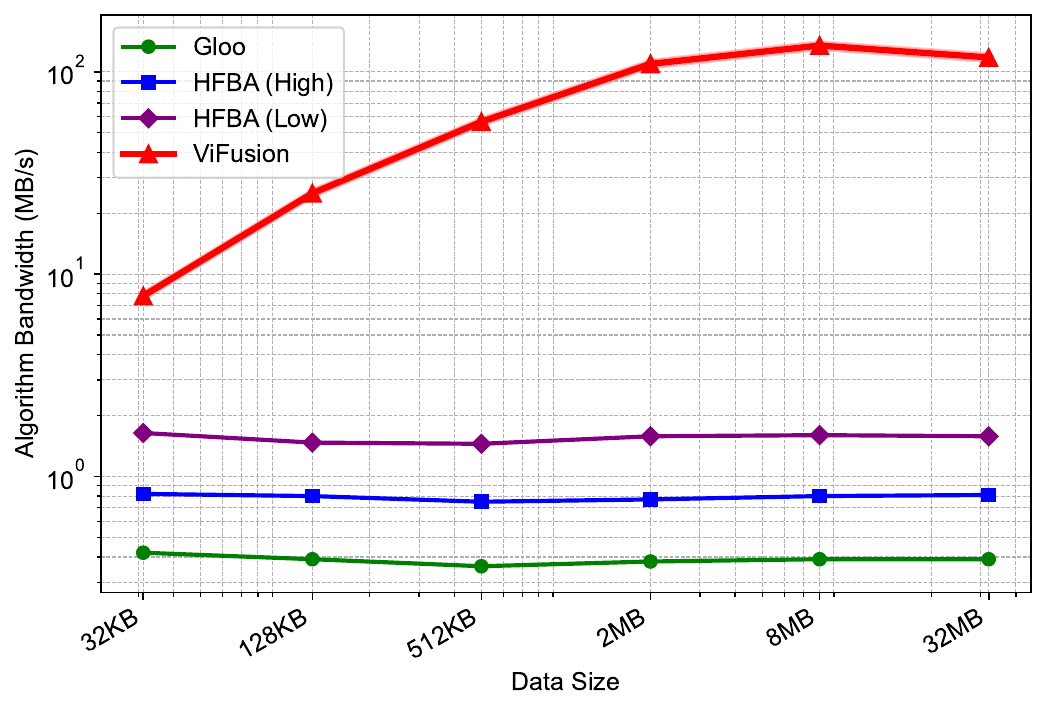}
    \caption{Comparison of the bandwidth of AllReduce in 8 GPU workers}
    \label{fig:8gpu_allreduce_ina}
  \end{subfigure}
  \caption{Overall comparison of AllReduce bandwidth}
  \label{fig:performance}
\end{figure*}

\subsection{Implementation}

To validate ViFusion's effectiveness, we conducted experiments on an 8-GPU server equipped with NVIDIA RTX 4090 GPUs as well as on multi-node CPU clusters for communication benchmarking. The implementation supports high-throughput tensor fusion with minimal overhead, while ensuring low-latency inference.

\subsection{Experimental Setup}

Our experimental setup consists of the following components.

\textbf{Hardware.} Our experimental setup consists of a high-performance computing environment designed to evaluate the efficiency and scalability of ViFusion. Specifically, we utilize a dedicated server equipped with eight NVIDIA RTX 4090 GPUs, interconnected via 40 Gbps high-speed links, ensuring efficient intra-server communication for parallel processing.

\textbf{Compared Framework.} To evaluate the effectiveness of ViFusion’s tensor fusion framework, we compare it against Gloo \cite{gloo}, a widely used open-source collective communication library optimized for distributed deep learning and heterogeneous computing environments.

\textbf{Baseline Method.} We evaluate ViFusion against multiple baseline methods, including Gloo \cite{gloo} and our locally implemented Hierarchical Fusion-Based Aggregation (HFBA). Gloo is a widely adopted, distributed communication library.

To further assess the effectiveness of ViFusion’s in-network aggregation, we introduce HFBA, a hierarchical communication approach that performs local reductions before global synchronization. HFBA follows a traditional multistage aggregation strategy that mimics the commonly used hierarchical AllReduce schemes. By omitting in-network computation optimizations, HFBA serves as a direct comparison point to evaluate how ViFusion leverages programmable switches and batch-based fusion to improve communication efficiency and throughput.

\subsection{Evaluation on Real-World Video Retrieval Tasks}

In order to thoroughly evaluate the end-to-end performance of the ViFusion framework, we developed a video retrieval system grounded in transformer-based models to investigate the feasibility of integrating text and video embeddings for large-scale video retrieval tasks. Specifically, we employ the CLIP model \cite{clip} to generate embeddings from both video frames and their corresponding textual descriptions. These embeddings are then stored in the Milvus \cite{Milvus} vector database, which provides robust scalability and accelerated similarity search operations. 

For our experiments, we leveraged a subset of the MSRVTT dataset \cite{msrvtt}, a standard benchmark often used in video understanding tasks. Each video in the chosen subset is paired with a textual description, providing a direct link between visual content and linguistic information. After extracting frame-level features via CLIP, we convert these features into vector embeddings and upload them to the Milvus database. Textual descriptions undergo a similar embedding process, resulting in embeddings that reside in the same vector space as the video frames. During retrieval, a user-input text query is similarly encoded, and a cosine similarity metric is used to identify the most relevant video embeddings within the database. We then retrieve the top ten matching videos based on their similarity scores, showcasing the system’s ability to semantically align textual queries with the corresponding visual content.

While we have not yet conducted comparative analyses against other established retrieval methods, the principal goal at this stage is to demonstrate the viability and efficiency of integrating transformer-based embeddings with the ViFusion framework. We place particular emphasis on evaluating how well the system can manage high-dimensional vector representations and facilitate real-time retrieval tasks when working with substantial amounts of data. Our initial findings are promising, with top-1, top-5, and top-10 mean hit ratios of 42.1\%, 71.2\%, and 81.3\%, respectively. These results point to the strong potential of the ViFusion-based approach for text-to-video retrieval, highlighting the capacity of transformer models to capture rich semantic information and of vector databases like Milvus to efficiently store and query these high-dimensional embeddings.

% For evaluating ViFusion’s end-to-end performance, we built a video retrieval system based on transformer models to assess the feasibility of integrating text-video embedding for large-scale video retrieval tasks. The system utilizes the CLIP model \cite{clip} to generate embeddings for both video frames and textual descriptions and stores them in a Milvus \cite{Milvus} vector database for fast similarity searches.

% The retrieval system operates on a subset of the MSRVTT dataset \cite{msrvtt}, in which each video is paired with a textual description. We performed a sample retrieval task in which a given text query retrieved the top ten most relevant videos based on cosine similarity.

% Despite not comparing the performance with baseline retrieval methods, the focus of this experiment is to demonstrate the effectiveness of integrating transformer-based embeddings with the ViFusion framework, highlighting its ability to efficiently manage high-dimensional vector data and perform real-time retrieval tasks. The results from this retrieval system show promising retrieval accuracy, with top-1, top-5, and top-10 mean hit ratios of 42.1\%, 71.2\%, and 81.3\%, respectively.

\subsection{In-Network AllReduce Performance}

Our experiments evaluate both throughput and end-to-end communication latency under different tensor fusion strategies, analyzing the impact of batch size, fusion granularity, and communication overhead on the overall performance of ViFusion.

To validate our in-network aggregation design, we conduct micro-benchmark tests using four and eight GPU workers, measuring the AllReduce bandwidth across a range of tensor sizes (from 32 KB to 32 MB). As a baseline, we compare ViFusion with Gloo and two hierarchical fusion-based aggregation (HFBA) strategies. HFBA (High) performs AllReduce on a larger number of smaller tensors (1 KB each), which increases the communication overhead owing to frequent synchronization. By contrast, HFBA (Low) aggregates fewer but larger tensors (4 KB each), reducing synchronization costs while still being constrained by host-based summation.

Figures~\ref{fig:4gpu_allreduce_ina} and \ref{fig:8gpu_allreduce_ina} show the average measured bandwidth for each method. For larger tensors (above 128~KB), the 4-GPU experiments reveal that ViFusion achieves approximately 37× to 43× higher throughput than Gloo. In addition, compared with HFBA, ViFusion improves throughput by approximately 8×–22× over HFBA (High) and approximately 5×–10× over HFBA (Low).

Although HFBA benefits from its hierarchical fusion strategy and exhibits better performance than Gloo, it remains bandwidth-limited, particularly at larger tensor sizes where ViFusion continues to scale effectively. HFBA (High) suffers from excessive communication overhead owing to frequent aggregations, whereas HFBA (Low) achieves a more stable performance but still struggles with host-based reductions.

The advantage of ViFusion is primarily attributed to its in-network computation approach, which significantly reduces the cross-rack communication volume.

By offloading computations to the network, our in-network approach alleviates congestion in uplink channels and eliminates redundant host-based summation operations. As the cluster size increases, these benefits become increasingly pronounced, demonstrating the critical role of in-network computation in large-scale distributed environments.

\begin{figure}[!t]
  \centering
  \includegraphics[scale=0.35]{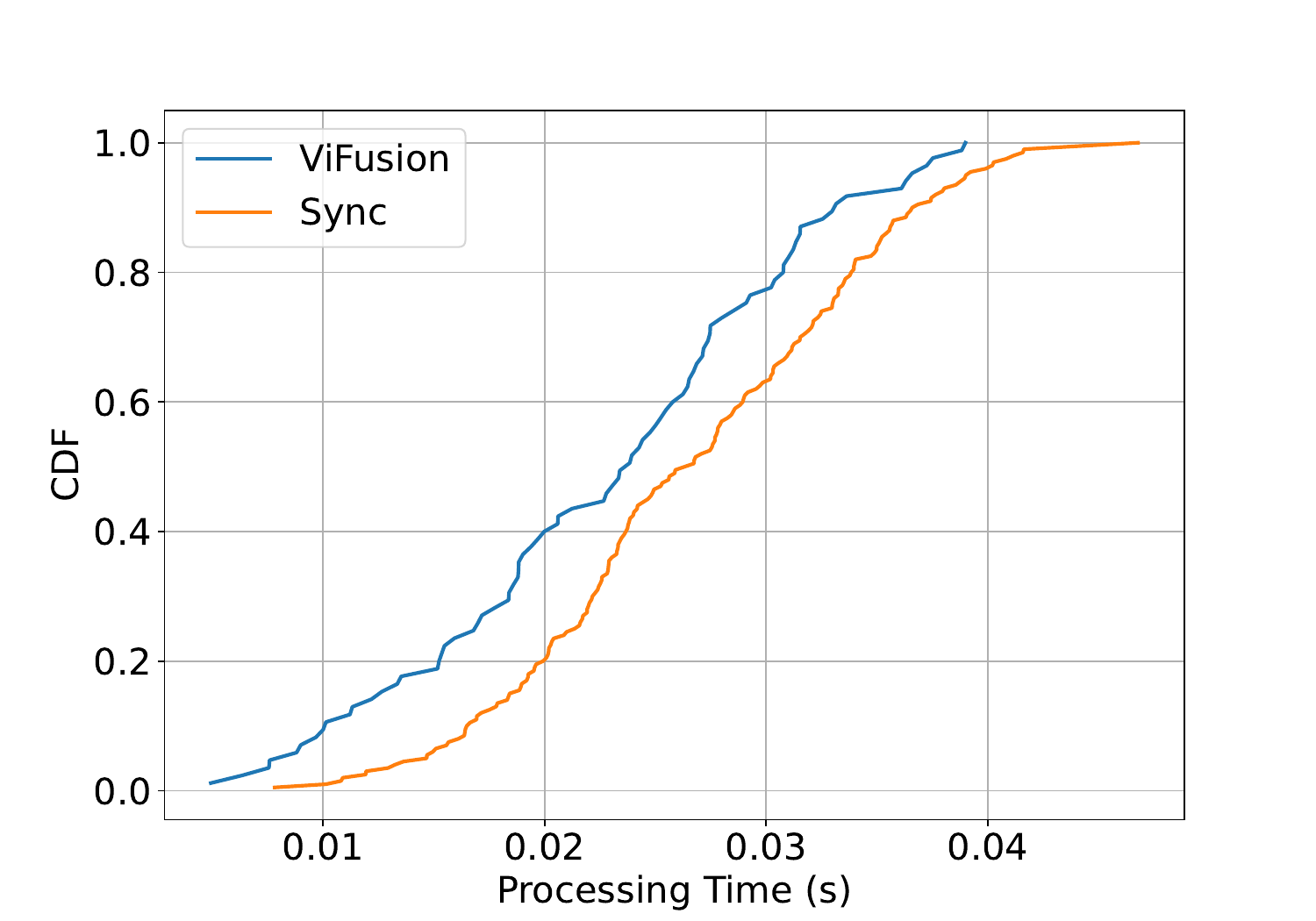}
  \caption{Comparison of Query Latency Distribution: ViFusion vs. Synchronous Baseline (Sync)}
  \label{fig:cdf}
\end{figure}

\subsection{Tensor Fusion Performance}

We benchmark ViFusion against a synchronous baseline (Sync), which processes queries individually without fusion, leading to an increased synchronization overhead. In contrast, ViFusion dynamically batches incoming queries using a dual-threshold mechanism, triggering fusion based on either a predefined batch size or timeout threshold.

The experiments are performed on an 8-GPU system, where each GPU is responsible for a shard of the model. Queries are generated asynchronously to simulate real-world workloads with interarrival times following a Poisson distribution. The primary metric evaluated is the latency distribution, which analyzes the end-to-end processing time of the queries.

Figure \ref{fig:cdf} compares the cumulative distribution function (CDF) of the query latency between ViFusion and Sync. The ViFusion curve is consistently left-shifted, indicating a lower latency across the entire distribution. At the median (50th percentile), ViFusion reduces the query latency by approximately 13\% compared with Sync. Furthermore, in the 99th percentile, ViFusion achieves a 23\% reduction in latency, highlighting its robustness under high-load conditions. These results demonstrate that fusion-aware scheduling effectively minimizes query completion time while maintaining stability under varying workloads.

\section{Conclusion}

In this paper, we present ViFusion, a communication-aware tensor fusion framework that addresses critical bottlenecks in large-scale video feature indexing. By unifying multilevel model sharding, zero-copy GPU data movement, and hierarchical AllReduce offloaded to in-network computing resources, ViFusion effectively reduces inter-node communication overhead and significantly increases GPU utilization. Our key innovations include dynamic batch fusion and networked aggregation. Experimental results on both GPU clusters and multinode CPU setups show that ViFusion not only achieves up to 8--22 times higher throughput compared to existing solutions, but also maintains comparable or lower end-to-end response times.

ViFusion’s flexible APIs and scheduler-based design make it readily adaptable to diverse neural architectures and deployment scenarios. Our hierarchical AllReduce scheme aligns well with typical data center topologies, alleviating cross-rack congestion and reducing straggler effects. Meanwhile, the tensor fusion module effectively aggregates many small video segments or feature tensors, substantially decreasing the kernel invocation overhead and idle time of deep learning accelerators.

\section{Acknowledgments}
\sloppy{
Guangdong provincial project 2023QN10X048, Guangzhou Municipal Key Laboratory on Future Networked Systems (2024A03J0623),
%the Guangdong Provincial Key Lab of Integrated Communication, Sensing and Computation for Ubiquitous Internet of Things (No.2023B1212010007),
the Guangzhou Municipal Science and Technology Project 2023A03J0011, the Guangdong provincial project 2023ZT10X009, and the
Natural Science Foundation of China (U23A20339). 
}

\bibliographystyle{ACM-Reference-Format}
\bibliography{main}

\appendix

% \subsection{Training Settings}









\end{document}